\documentclass{article}

\usepackage{arxiv}

\usepackage[utf8]{inputenc} % allow utf-8 input
\usepackage[T1]{fontenc}    % use 8-bit T1 fonts
\usepackage{hyperref}       % hyperlinks
\usepackage{url}            % simple URL typesetting
\usepackage{booktabs}       % professional-quality tables
\usepackage{amsfonts}       % blackboard math symbols
\usepackage{nicefrac}       % compact symbols for 1/2, etc.
\usepackage{microtype}      % microtypography

\usepackage{lipsum}         % Can be removed after putting your text content
\usepackage{graphicx}
\usepackage{natbib}
\usepackage{doi}

\usepackage{times}
\usepackage{latexsym}
\usepackage{amsmath}
\usepackage{cleveref}       % smart cross-referencing
\usepackage{amssymb}
\usepackage{booktabs} 
\usepackage{paralist}
\usepackage{subcaption}
\usepackage{comment}
\usepackage{afterpage}
\usepackage[T1]{fontenc}
\usepackage[utf8]{inputenc}

\usepackage{microtype}

\usepackage{inconsolata}

\usepackage{tabularx} 
\usepackage{graphicx}
\usepackage{xcolor}

\title{Training Table Question Answering via SQL Query Decomposition}

\author{ 
    {\includegraphics[scale=0.06]{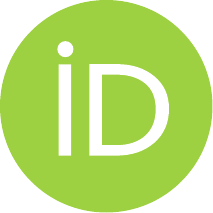}\hspace{1mm}Raphaël Mouravieff} \\
    Sorbonne University\\
    MLIA, ISIR\\
    F-75005 Paris, France \\
    \texttt{raphael.gervillie@isir.upmc.fr} \\
    \And
    {\includegraphics[scale=0.06]{orcid.pdf}\hspace{1mm}Benjamin Piwowarski} \\
    Sorbonne University\\
    CNRS, MLIA, ISIR \\
    F-75005 Paris, France \\
    \texttt{benjamin.piwowarski@isir.upmc.fr} \\
    \And
    {\includegraphics[scale=0.06]{orcid.pdf}\hspace{1mm}Sylvain Lamprier} \\
    University of Angers\\
    LERIA, SFR MATHSTIC\\
     F-49000 Angers, France \\
    \texttt{sylvain.lamprier@univ-angers.fr} \\
}

\renewcommand{\headeright}{Preprint}

\hypersetup{
pdftitle={Learning Relational Decomposition of Queries for Question Answering from Tables},
pdfsubject={cs-AI},
pdfauthor={Raphaël Mouravieff},
pdfkeywords={First keyword, Second keyword, More},
}

\begin{document}
\maketitle

\begin{abstract}

% Studies in table question answering have highlighted the benefits of intermediate pre-training from SQL queries. This breaks the complexity of the task, which 
Table Question-Answering involves both understanding the natural language query and grounding it in the context of the input table to extract the relevant information. 
In this context, many methods have highlighted the benefits of intermediate pre-training from SQL queries. 
However, while most approaches aim at 
generating final answers from inputs directly, 
we claim that there is better to do with  SQL queries during training.
By learning to imitate a restricted portion of SQL-like algebraic operations, we show that their execution flow provides intermediate supervision steps that allow increased generalization and structural reasoning compared with classical approaches of the field. 
Our study bridges the gap between semantic parsing and direct answering methods and provides useful insights regarding what types of operations should be predicted by a generative architecture or be preferably executed by an external algorithm.
 
\end{abstract}

% keywords can be removed

\section{Introduction}

The field of Table Question Answering (QA), which encompasses complex content manipulation tasks like projection, sorting, grouping, and aggregation, presents considerable challenges for Natural Language Processing (NLP). Its complexity and growing relevance across diverse sectors, from business to academic research, have attracted widespread attention. This domain has evolved quickly with the rise of Pretrained Language Models (PLMs), but this field remains challenging for current models~\cite{jin2022survey}. 

Former studies focused on Semantic Parsing (SP) techniques tailored for well-structured and clean table data, as highlighted in \cite{shi2020potential}. However, real-world scenarios often involve heterogeneous resources, for example combining both numerical and textual content in some cells, like in WikiTableQuestions \cite{pasupat2015compositional}. Among the proposed solutions,  \cite{liu2021tapex} tried to \emph{generate directly} the answer and therefore bypass the generation of logical forms. Despite this advantage, these methods exhibit limitations, particularly when executing numerical operations (e.g. computing a mean, counting). 
To cope with this, a natural solution is to propose hybrids that stand as intermediary solutions between semantic parsing and direct generation. For instance, \cite{herzig2020tapas,zhou2022unirpg} have combined basic table selection methods (e.g. selecting rows and columns, or cells) before computing aggregations or performing basic numerical operations. However, they often fail to address intricate queries necessitating the synthesis of diverse table views and interactions because of the limited expressivity of their underlying algebra.
% --- The model
In this work, we propose to study the continuum between semantic parsing-based and direct generative methods, to leverage the strengths of both. Going beyond previous works, we propose a novel framework that facilitates reasoning over heterogeneous table resources. This framework relies on the definition of an algebra over tables inspired by relational algebra. Based on this algebra, each question in natural language and its corresponding table can be translated into a computational graph. By varying a cut-off criterion that specifies which part of the graph should be computed directly by the model (i.e. direct generation) and which one should be computed outside of it (i.e. semantic parsing), we can study different trade-offs and their effect in terms of effectiveness.

 Beyond a stronger interpretation of the user query in the context of the table compared to semantic parsing-based (SP) methods, our framework addresses the common execution challenges associated with SP methods, %we introduce a 'table cleaning' phase during execution, enhancing the adaptability of our logical form for future operations and ensuring its compatibility with real-world tables. 
 which require clean tables to allow full SQL  execution.
 Our approach predicts operators with associated  "clean" operands from the input, 
 thanks to the generation ability of the Transformer architectures. 

To learn our model, we leverage a pre-training procedure~\cite{pruksachatkun2020intermediate,geva2020injecting,yu2020grappa} that helps neural architectures to manipulate tabular data, before dealing with complex Table QA tasks,  by first learning to generate from SQL queries rather than from natural language. We then perform experiments showing that our model performs as well as state-of-the-art models relying on much more sophisticated training procedures. More importantly, we show that for some intermediate cut-off levels, our approach allows us to better generalize and is more robust compared to direct answer methods, which are usually limited in their structural reasoning capacities.

\section{Related Work}

\begin{figure*}[tp]
  \centering
  \includegraphics[width=\linewidth]{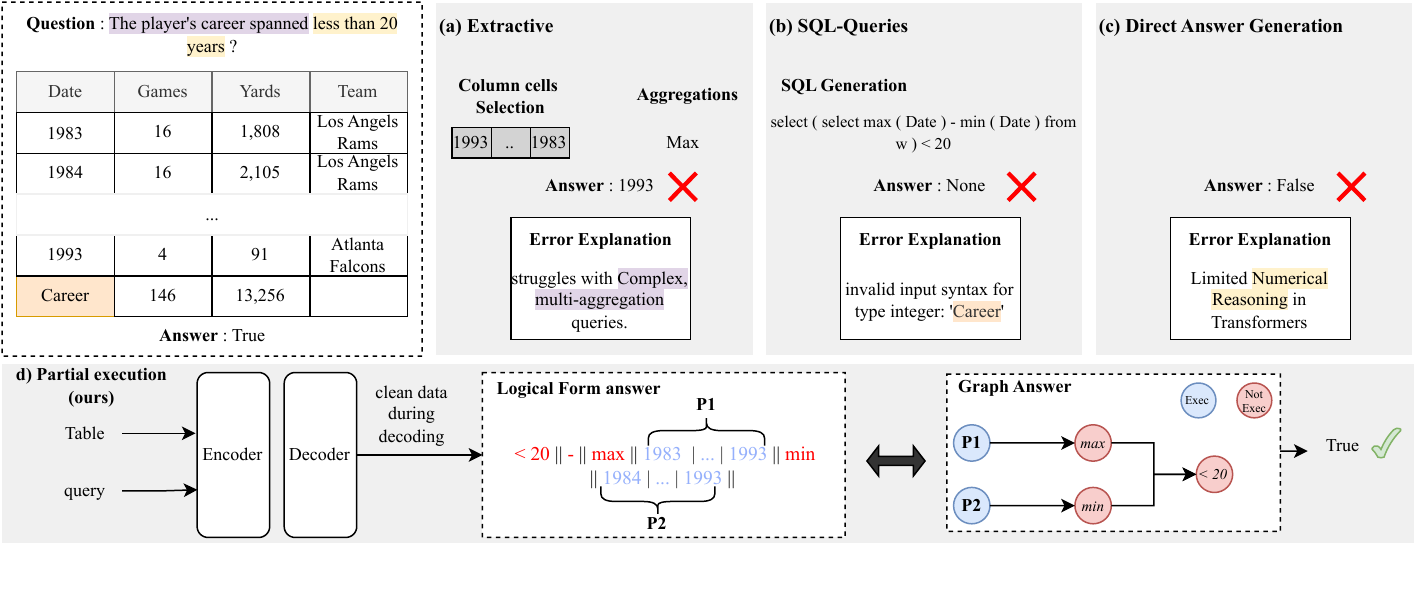}
  \caption{Overview of the different approaches for Table QA and their limits (a-c), along with our proposition (d)}
  \label{fig:model}
\end{figure*}

\subsection{Table Question Answering Architectures}

Table question answering is a very active field with many recent developments. 
This ranges from specifically designed transformer architectures, with sparse \cite{eisenschlos2021mate} or biased \cite{golchin2023time}  attention matrices that capture table structures, or specialized table embeddings as in TUTA~\cite{wang2021tuta}, %enhance the model's ability to understand table structure
to large Language Modelds (LLMs) that leverage in-context learning to deal with table structures~\cite{chen2022large,cheng2022binding,wang2024chain}.
While our study,  orthogonal to these directions, could be applied in the context of any family of architectures including LLMs, e.g. fine-tuned using low-rank adaptation \cite{hu2021lora,dettmers2024qlora}, we chose to build on compact architectures, based on reasonably-sized pre-trained language models (PLMs) such as BERT or BART, as considered in popular recent works TAPEX~\cite{liu2021tapex} or OmniTab~\cite{jiang2022omnitab}. 
Beyond scalability, such architectures, which do not require specific prompt design that could bias conclusions, offer easier comparison opportunities.\footnote{We also note that it has recently been shown in a broader context that LLMs are usually contaminated by evaluation benchmarks \cite{golchin2023time}, which could alter the results of our study.} 
Finally, we believe that our developed approach, which consists of predicting and using external programs as tools when generating answers, e.g. in the vein of Toolformer~\cite{schick2024toolformer}, are still fully valuable in the context of LLMs, providing increased inference speed and stability. Our work is a step into showing how tools can be used with structured data like tables, which can be transferred to LLMs in future works.

In the following, we focus on differences between table question answering approaches regarding their output strategies, which is more strongly related to the study of this paper. %We review these  

\subsection{Output Strategies in Table QA}

Table Question-Answering models can be distinguished on their answer generation, which is either a formula operating on the table (semantic parsing) or a direct answer (direct generation), or a hybrid of both.

\paragraph{Semantic Parsing} Semantic parsing aims to transform natural language into executable queries, primarily SQL. Sketch-based models decomposed SQL query construction by breaking down and classifying query components, enhancing structured SQL generation~\cite{jin2022survey}. Generation-based methods like RAT-SQL~\cite{wang2019rat} directly produce SQL queries using an encoder-decoder architecture that considers both the question and the table context for generation. Under weak supervision, \cite{min2019discrete} optimize the probability of the correct answers over a set of possible latent representations,

facilitating the model's ability to infer correct responses without explicit answer mappings. 
Another stategy is to use reinforcement learning where the execution result is used as rewards to train models \cite{zhong2017seq2sql}. Despite SQL's effectiveness in QA over tables \cite{shi2020potential}, its limitations with non-database tables and question translation are a major drawback. Our approach seeks to transcend these bounds by introducing a logical form independent of the table during execution.

\paragraph{Direct Answer Generation} In contrast to semantic parsing, direct answer generation produces final answers, bypassing the step of converting questions into formulas. This directly addresses the limitations of SQL-based systems, enabling the processing of various table formats. For instance, \cite{mueller2019answering} use a GNN-based encoder to encode the table structure and a decoder to output the answers conditioned on the graph and the query. An additional benefit of this method is its compatibility with advanced data augmentation techniques~\cite{eisenschlos2020understanding}. This includes transformations from SQL to its result as in TAPEX~\cite{liu2021tapex}, or from Excel formula to its execution as in FORTAP~\cite{cheng2021fortap}. However, a notable challenge for transformers in this domain is handling numerical reasoning queries effectively~\cite{zhou2022tacube}. 

\paragraph{Hybrid Methods}
Hybrid methods extract pertinent tokens from tables to create responses, typically employing an aggregator to associate with and route these tokens to a specifically designed executor.  TAGOP~\cite{zhu2021tat}  uses sequence tagging for extracting relevant cells and a classifier for assembling them into coherent symbolic reasoning programs. TAPAS \cite{herzig2020tapas} employs a classifier layer at the end of a BERT-like encoder for selecting content from tables and determining the aggregation operation to apply to it. These methods have good numerical abilities, but however, unlike other output strategies, they have limited expressiveness and struggle with complex multi-aggregation queries \cite{herzig2020tapas}. Our proposed supervision using intermediate logical form addresses this issue by enabling complex multi-aggregation representations.

\section{Model}

The goal of Table QA is to find the answer $A$ given a natural language question $q$ posed on a table $T$. In this section, we first describe the algebra that we use to represent an SQL query. We then describe how to translate formulas using this algebra into different sequences that depend among other things on the desired level of granularity.

\begin{table*}
\centering
\begin{tabular}{llll}
\hline
\textbf{Operation} & \textbf{Function Definition} & \textbf{Parameters} & \textbf{Description}\\
\hline
% P
Projection  & $P: {\cal T}  \rightarrow {\cal T}$ & $J=\{c_i\}_{i\in 1\ldots k}$ & Extracts $k$ columns from a table T, \\ & & & specified by their names $J \subseteq {h_T}$.\\\hline\hline
% +C
Comparison  & $C: {\cal T} \cup {\cal G} \times {\cal T}  \rightarrow {\cal B}$  &  $c\in \{>, <,..\}$ & Compares $T_1 \in {\cal T} \cup {\cal G}$ with $T_2 \in {\cal T}$ using  $c$.\\
& & & $T_2$ either has the same number of rows as $T_1$ \\
& & & or only 1 that is broadcast to fit $T_1$. \\\hline\hline
%+GB+H
Having  & $H: {\cal G} \times  B \rightarrow {\cal G}$ & - & Selects from $G$ where $B$ is true,\\ & & & with $N^B_{rows}=N^G_{rows}$. \\\hline
Group By & $GB: {\cal T}  \rightarrow G$ & $J=\{c_i\}_{i\in 1\ldots k}$  & Groups elements in $T$ with equal \\ 
& & & values from columns in $J\subseteq h_T$.\\\hline\hline
% + A
Aggregation & $A: {\cal T} \cup {\cal G} \rightarrow {\cal T}$  & $f \in \{sum, avg,..\}$ & Aggregates $T$ using function $f$.\\\hline\hline
% + OP
Operator & $OP: {\cal T} \times {\cal T}  \rightarrow {\cal T}$ & $o \in \{+,-,*,..\}$ & Performs the term-wise operation $o$ on \\ 
& & & two tables $T_1 \in {\cal T}$ and $ T_2 \in {\cal T}$.\\\hline\hline
% Remains
Order By  &  $OB: {\cal T}  \rightarrow {\cal T}$ &  $d \in \{asc, desc\}$ & Orders table $T$ by criterion with direction $d$.\\\hline
Limit & $L: {\cal T} \rightarrow {\cal T}$ & $k\in \mathbb{N}$ & Selects top $k$ elements from $T$.\\\hline
Selection  & $S: {\cal T} \times B \rightarrow {\cal T}$ & - & Selects from $T$ where $B$ is true, \\ & & & with $N^B_{rows}=N^T_{rows}$. %owns $N^T_{rows}$ values. 
\\\hline
\end{tabular}
\caption{\label{table:operations} Algebra to manipulate tabular data. See section \ref{sec:algebra} for notations.}
\end{table*}

%Our propose met

\subsection{Tabular algebra}
\label{sec:algebra}

% In this section, we describe the algebra that we use to manipulate tabular information. 
% Given that queries in natural language can be transformed into SQL queries, 
In this section, we describe the algebra, inspired by the relational one \cite{codd1970relational}, 
that we use to represent any operation on tables.

\paragraph{Structures}
Table Question Answering is the task of finding an answer $A$ from a table $T \in \mathcal{T}$, where % is defined as a 
$T = \left( (x_{r,c})_{c=1\ldots N^T_{col}} \right)_{r=1\ldots N^T_{row}}$ is a matrix of  values $x_{r,c}$, which can be numbers or strings. 
% Classically, tables only include atomic values.  
Differently from relational algebra, we view tabular data as a sequence of tuples which we suppose to be \emph{ordered}.
%To cope with set aggregations (i.e., involving a group by operation), %with an operator),
%we also consider $x$ values of set of values. %manipulate group-by tables:
A table can have a header, which corresponds to a sequence of column names $c_1 \ldots c_{N^T_{col}}$. When no header is given, each $c_i$ corresponds to the column index, and $h_T=\{c_i\}_{i=1}^{N^T_{col}}$  stands as the set of column names from $T$.  %We  correspond to %also have
%names associated with each column $c$.
% This work does not consider tables with merged columns on some rows, or in higher dimensions than 2,  but it could be  extended to such settings easily, providing specific operators. 
Views on the original table, that correspond to results from algebraic operations, are also considered as table $T \in {\cal T}$. 

Classically, tables only include atomic values.  
To cope with set aggregations (i.e., involving a group-by operation), %with an operator),
we also manipulate group-by tables $G \in {\cal G}$, where 
$G = \left( (g_{r,c})_{c=1\ldots N^G_{col}} \right)_{r=1\ldots N^G_{row}}$, with each component $g_{r,c}$ corresponding to a set of values. 
%
%$$G = \left( (x_{r,1}, \ldots, x_{r,N_{col}}, T_r) \right)_{r=1\ldots N_{row}}$$
%where the table $T_r$ corresponds to a part of the table associated with the tuple $(x_{r,1}, \ldots, x_{r,N_{col}})$.
%
We also note columns boolean matrices %consider a subset of
%as $B$ boolean vectors tables that contain only boolean values as 
as $B =  \left(b_{r,1}) \right)_{r=1\ldots N^B_{row}}$, 
with $b_{r,1}\in \{0, 1\}$.

% Let $T = \{x_{r,c}|1\leq r \leq N_{row}, 1\leq c \leq N_{col}\}$ be a table with $N_{row}$ rows and $N_{col}$ columns, and $x_{r,c}$ a cell of the table at row $r$ and column $c$. $R$  a relation corresponding to a view of $T$. Also,  $B = \{b_i|b_i \in \{0,1\}, 1\leq i \leq |B| \}$ a Boolean vector. Executing logical form $S$ on table $S(T)=A$ produce also the final answer.

\paragraph{Operators}

Table \ref{table:operations} describes the different operators that we use to manipulate tables $T$ or group-by tables $G$, %which operate on the above-defined data and
whose behavior can be conditioned on parameters (e.g. ``order by'' can be ascending or descending). These operators follow roughly standard relational algebra operators and cover a broad range of SQL queries. A notable difference with classical relational algebra, which was dictated by the fact we want to further decompose operations for analysis purposes, is the fact that the selection operation simply corresponds to a filter given a column of boolean values produced by a separated comparison operator and that the order of tuples is used for comparisons (e.g. >, <) and operations (e.g. +, -).    %There are two main differences with relational algebra, which were dictated by the fact that we wanted a simple token-based representation of any input and output of an operator:

%\begin{enumerate}
%    \item To cope with conditions (\texttt{WHERE}), we rely on two tables with the same number of rows: the first one contains the data to be filtered, and the second boolean values (i.e. $B$ with values in $\{0, 1\}$;
%    \item Group-by operators generate not tuples, but tuples associated with a table. 
%\end{enumerate}

\newcommand{\node}[1]{\phi\left( #1 \right)}
\newcommand{\nodes}{\mathbf{N}}

Translating from SQL to our algebra is straightforward. We rely on the SQLGlot library\footnote{\url{https://github.com/tobymao/sqlglot}} 
to obtain a parse tree from any SQL query. This parse tree is then translated into a computational graph.
Each node $n$ of this graph is denoted as $\node{x_n, [n_1, \ldots, n_K]}$ where $x_n$ is either a table in $\mathcal{T}$, a group-by table in $\mathcal{G}$ or an operator in $\mathcal{O}$ (an operator is both the operation, e.g. "limit", and its parameters, e.g. $k$). In the case of operators, $n_1,\ldots,n_K$ correspond to the arguments of the operators, i.e. other nodes in the computation graph corresponding to its operands, and $x_n(.)$ the application of the operator on the corresponding list of child nodes. % the inputs of the operator $x$.
By abuse of notation, in the following we note $n=\node{x_n, [n_1, \ldots, n_K]} \in {\cal X}$, with ${\cal X}$ a given set, to denote $x_n \in {\cal X}$.  

% , with $N$ as the set of operations $\mathcal{O}$, t  ables or group-by tables $\mathcal{T}$, and $E$ are the labeled edges between each node. Labels allow to distinguish the different operands for operators. Moreover, an operator $op$ can be associated with some arguments (e.g. ``asc'' for order-by). 

\subsection{Partial Execution of the computational graph}

Now that we have defined the data and the algebra, we can present how this can be leveraged to produce various representations. 
For this, we rely on a graph transduction function $v$ operating recursively on any node $n$ of the graph. That is, given a set of operators ${\cal O}^*$ we allow to be executed,  $v(n)=\phi(x_n( v(n_1), \ldots, v(n_K)))$  if  $x_n \in {\cal O}^* \wedge \forall i \in 1\ldots K, v(n_i)\in \mathcal{T} \cup \mathcal{G} $, and $v(n)=n$ otherwise.   %. Given a criterion $\mathbf{C}(n)$:  
%The function \( v \) can transform each node of the graph in three different ways:
%\begin{itemize}
%    \item \textbf{Identity:} If the node $n$ is a terminal node, i.e. $x_n \in \mathcal{T}\cup\mathcal{G}$, we do not transform it;
%    \item \textbf{Execution:} If the current node $n$ corresponds to an operator that meets the criteria $\mathbf{C}(n)$ for being executed, the node is replaced by the result of its execution;
%    \item \textbf{Transformation:} If the node does not meet the criteria, we transform its arguments but keep the operator
%    as is, effectively modifying the graph's structure while preserving its logical flow.
%\end{itemize}
%
%Formally, this can be written:

In other words, we execute from any leaf to the root of the computational tree every allowed operation in ${\cal O}^*$ until execution is blocked (because $x_n$ not in ${\cal O}^*$ or one of its dependencies cannot be executed).

% More formally, whether a node is executed or not depends on several criteria. We define $\mathcal{O}^*_{excl} \subseteq \mathcal{O}^*$, the set of operators we do not wish to execute. $\mathcal{O}^*_{excl}$ acts as a hyperparameter that controls which part of the graph should or should not be executed. We also define $\Sigma_i \subseteq \Sigma$, the set of operands of the node  $n_i$. 

The computation graph %\( G \)
can hence be partially executed through this transformation $v$, 
allowing for flexible handling of SQL operations,  by applying $v$ on the root node.

\subsection{Linearizing the Graph}

As the computational graph must be generated sequentially, 
we need to define how to transform it into a sequence of tokens, i.e. how to \emph{linearize} it. 
To do so, we use a linearization function that we denote $l$, 
which takes a node $n= \node{ x_n, n_1, \ldots, n_K }$ in input and returns a sequence of tokens.

In the case of tables (i.e., when $x_n \in {\cal T}\cup{\cal G}$), we use a simple markup where we separate rows with the symbol ``\texttt{|}'' and columns with a comma ``\texttt{,}''. In the case of operators, i.e. $x_n\in\mathcal O$, the linearization $l$ corresponds to the name of the operator followed by its parameters. For instance, the sequence \texttt{LIMIT 1} corresponds to the limit operator with 1 as its parameter.

For operator nodes, we define various linearization schemes depending on the order (pre- or post-order) and the usage of aliases to avoid duplicating the same information (the graph is a directed acyclic graph, but there can be different paths between two nodes since results might be re-used).

\paragraph{Pre-order vs post-order}

We can either use a pre-order linearization scheme where the operator appears before its operands:
%\begin{multline*}
$l_{pre}(n)=  l(x_n) \oplus \oplus_i ( || \oplus l_{pre}(n_i))$ 
%\end{multline*}
or a post-order one:
%\begin{multline*}
$l_{post}(n)=  \oplus_i ( || \oplus l_{post}(n_i))  \oplus || \oplus l(x_n)
$. 
%\end{multline*}
In both cases, '$||$'denotes a separator token and $\oplus$ concatenation.

\paragraph{Using aliases}

In the above linearizations, re-used results will be linearized several times. This happens frequently with queries with some aggregation. The problem is that this can result in longer sequences, which in turn might be harder to generate. To tame this problem, we associate each node with a given alias the first time it is linearized (e.g. \texttt{N13}) and use this reference instead of its linearization in subsequent occurrences (see appendix \ref{sec:appendix:aliases} for details).

Finally, tables are linearized either before or after the operators. % (hence $\varnothing$ in the formula above). 
After some preliminary experiments, we chose this to make the grammar of the sequence more regular for a transformer (not mixing operators and content).

\section{Experiments}

\subsection{Dataset and Evaluation Metrics}

In our experiments, we used the WikiTableQuestions (WTQ) dataset~\cite{pasupat2015compositional}, the only dataset that fulfills all predefined criteria for our study: It is characterized by its provision of complex numerical reasoning questions, tables with missing information, mixed cell types (e.g. text and numbers), and availability of SQL supervision. The SQL annotations supplied by SQuALL~\cite{shi2020potential} enable the coverage of approximately 80\% of the questions from the WTQ, in the training and validation sets only.

Results are reported using the Denotation Accuracy (DA) metric as our primary evaluation criterion. DA checks if the execution of the predicted answer is equal to the target answer. When the answer is a list of results, DA disregards the order (i.e. set equality).  
% DA measures the percentage of predicted answer(s) that are equal to the ground-truth answer(s).
% Compared to Exact Match, DA deals with the case where an answer is a list, since DA disregards the list order.
We decomposed this metric into two categories: the Strict Denotation Accuracy (SDA), which is the traditional one used, and the Flexible Denotation Accuracy (FDA), which compares results after removing units (years, \$, kg, etc.). The choice to employ both SDA and FDA stems from our dependence on external tools' APIs for execution. As a result, our execution outcomes are unit-less, and using SDA would hide the improvements brought by our model -- note that we could extend our method to generate an arbitrary sentence containing the result in future works.
% Hence, we reassess models from the literature using both metrics, aiming to accurately showcase our models' performance on the WikiTableQuestions dataset.

\subsection{Inputs and outputs}

The query encoding is straightforward but table encoding presents a challenge due to its inherent structure. We follow TAPEX and OmniTAB \cite{liu2021tapex,jiang2022omnitab}, and represent the transformed table as $T^*$ = [HEAD], $c_1$, ..., $c_N$, [ROW], 1, $r_1$, [ROW], 2, $r_2$, ..., $r_M$. The tokens \texttt{[HEAD]} and \texttt{[ROW]} delimit the table's header and row sections, respectively, with subsequent numbers indicating row indices. Additionally, we use a vertical bar | to delineate headers or cells in separate columns. We then concatenate the query with the linearized table as the input of the encoder.

Outputs in our model correspond to linearized computational graphs. We considered 42 experimental conditions.
First, we use one of the following seven sets of operators as $\mathcal O^*$:
\begin{inparadesc}
    % \item[($\emptyset)$] No execution;
    \item[(P)] Only projection operators;
    \item[(+C)] P with comparison operators;
    \item[(+S)] +C with selection operators; 
    \item[(+GB+H)] +S with group-by and having;
    \item[(+A)] +GB+H with aggregations;
    \item[(+OP)] +A with operators;
    \item[(Full)] with all operators, i.e. as TAPEX~\cite{liu2021tapex}.
\end{inparadesc}
Second, we used six possible linearizations: pre-order, post-order, and pre/post-order-alias-start/end. Examples of different linearizations, with different partial executions, are given in the appendix \ref{sec:appendix:linearization}, tables \ref{tab:example:preorder}, \ref{tab:example:postorder}, \ref{tab:example:preorder-alias} and \ref{tab:example:postorder-alias}.

\subsection{Training pipeline}

Our training methodology employs a standard sequence-to-sequence (seq-2-seq) framework, with BART as the backbone architecture ~\cite{lewis2019bart}. 
We use the TAPEX~\cite{liu2021tapex} checkpoint to initialize our parameters and follow the proposed pre-training procedure, as preliminary experiments have shown improved results. Following TAPEX~\cite{liu2021tapex}, this process is divided into two distinct stages where we maximize the likelihood of the linearized relational formula (section \ref{sec:algebra}):
%\begin{inparaitem}[(i)]
    %\item 
    (i) We pre-train the model to translate SQL queries into our logical form. This step is crucial for adapting the model to understand the structure and semantics of SQL queries in the context of our logical representation;
    (ii) we fine-tune our model using natural language questions instead of SQL. Our additional hyper-parameters only correspond the choice of operators in ${\cal O}^*$ from the  validation set, as discussed below. 
%\end{inparaitem}

\subsection{Overall performance}

In this section, we compare our model with the state-of-the-art ones, on the test split of the WTQ Dataset. Results are shown in Table \ref{tab:overall}, distinguishing between those employing fine-tuning techniques from BART-like architectures and those considering in-context learning of LLMs, using %utilizing 
specific prompting strategies. We report SDA as well as FDA for the model for which we reproduced the results.
We report in table \ref{tab:overall} the results of the best-performing set of operators we experimented, namely  $\mathcal O^*=\{P,C,S\}$, as well as our ensemble model that leverages various granularities $\mathcal{O}^*$.

We can first note that prompting approaches based on LLMs, including the cutting-edge chain-of-thought method~\cite{wei2022chain}, demonstrate superior performance without necessitating model adaptation.
At the other end of the spectrum, the semantic parsing baseline SQuALL does not perform well, especially if tables are not manually cleaned up (dropping from 54.3 to 27.2 for FDA), while other methods do not require this costly cleaning step.
Our models showcase notable achievements, with our best one (selected on the validation set) reaching an FDA of 61.4\%. This is comparable to OmniTab which relies on sophisticated data augmentation techniques. We can even increase to 66.3\% when leveraging ensemble methods (see section \ref{sec:ensembling}). We also show later that besides obtaining state-of-the-art results (for similarly sized architectures), our models are also more robust.

\begin{table}[t]
\centering
\caption{Comparison of Model Performance}
\begin{tabular}{lcc}
\hline
\textbf{Model} & \textbf{SDA} & \textbf{FDA} \\
\hline
\multicolumn{3}{c}{\textit{Fine-Tuned BART-like Models}} \\

TABERT \cite{yin2020tabert} & 52.3 & - \\
MATE \cite{eisenschlos2021mate} & 51.5 & - \\
TableFormer \cite{yang2022tableformer} & 52.6 & - \\
GRAPPA \cite{yu2020grappa} & 52.7 & - \\
DoT \cite{krichene2021dot} & 54.0 & - \\

REASTAP \cite{zhao2022reastap} & 58.6 & - \\
TaCube \cite{zhou2022tacube} & 60.8 & - \\

TAPAS \cite{herzig2020tapas} & 48.8 & 50.2 \\
TAPEX \cite{liu2021tapex} & 55.5 & 57.9 \\
OmniTab \cite{jiang2022omnitab} & 61.8 & 62.1 \\

\multicolumn{3}{c}{\textit{Prompt-based LLMs}} \\
ChatGPT \cite{cheng2022binding} & 43.3 & - \\
Codex \cite{ye2023large} & 47.6 & - \\
StructGPT \cite{cheng2022binding} & 48.4 & - \\
Codex-COT \cite{chen2022large} & 48.8 & - \\
Binder \cite{cheng2022binding} & 64.6 & - \\
LEVER \cite{cheng2022binding} & 65.8 & - \\
DATER \cite{cheng2022binding} & 65.9 & - \\

Chain-of-Table \cite{wang2024chain} & 67.3 & - \\
\multicolumn{3}{c}{\textit{Semantic parsing on test with cleaned tables}} \\
SQuALL \cite{shi2020potential}  & 50.4 & 54.3 \\
\multicolumn{3}{c}{\textit{Semantic parsing on test tables}} \\
SQuALL \cite{shi2020potential}  & 23.2 & 27.2 \\

\multicolumn{3}{c}{\textit{Our models}} \\
% +P  & 47.5 & 50.1 \\
% +C  & 54.7 & 57.7 \\
+P+C+S  & 59.0 & 61.4 \\
% +GB+H  & 58.6 & 61.2 \\
% pas la peine finalement, cf mattermost
% +A  &  &  \\
% +OP  &  &  \\
% Full &  &  \\
Ensemble  & 63.3 & 66.3 \\
\hline
\end{tabular}
\label{tab:overall}
\end{table}

% \subsection{Effects of pre-training on fine-tuning performance}

% TODO TODO TODO

% \begin{table*}[htbp]
% \centering
% \label{tab:sensitivity:op}
% \caption{Comprehensive Comparison of Configurations Across Preprocessing Techniques. Configurations (P, PC, PCS, PCSGHB, PCSGHBOB, PCSGHBOBA, PCSGHBOBAOP, Full) are derived using the preorder\_alias\_end linearization. Rows correspond to different pretraining methods, and columns to the fine-tuned model.}
% % \resizebox{\textwidth}{!}{%
% \begin{tabular}{lccccccccc}
% \hline
%   \textbf{Pre-training}  & \textbf{+S} & \textbf{+GB+H} & \textbf{+OB} & \textbf{+A} & \textbf{+OP} & \textbf{Full Exec} \\
% \hline
% No  & 54.0 & 55.2 & 55.1 & 42.8 & 42.3 & 39.5 \\
% Tapex  & 58.2 & 58.1 & 58.5 & 57.1 & 57.5 & 57.9 \\
% +S  & 57.1 & 58.6 & 58.3 & - & - & - \\
% +GB+H  & - & - & - & - & - & - \\
% +OB  & - & - & - & - & - & - \\
% +A  & - & - & - & - & - & - \\
% +OP  & - & - & - & - & - & - \\

% \hline
% \end{tabular}%
% % }
% \end{table*}

\subsection{Sensitivity over questions types}
\label{sec:types}

In table \ref{tab:sensitivity:op}, we show the performance for different query types, distinguished by whether they contain operators such as Projection, Comparison, Selection, Group By, Order By, Aggregation, Operator, and Limit. Note that queries containing a group-by are limited (30), and hence results reported in this column should be taken with care.
% To compute the means, we weigh the FDA of each query by the number of times the operator appears in the computational graph. 
% Doing so allows to measure the impact of each type of operator depending on its prevalence.
% of different models -- Omnitab, Tapex, Tapas, SQuALL, and ours -- perform .
% A closer examination reveals nuanced insights into the strengths and weaknesses of each model. 

Among existing models, Omnitab has the strongest performance, showing the importance of its data augmentation techniques compared to Tapex, especially for complex operators such as \emph{group by} and \emph{operators}. Tapas does perform  worse on these query types, which shows the limits of its aggregation methodology based on column/row selection.

% Omnitab consistently exhibits strong performance across the board, particularly excelling in handling Order By and Limit clauses, suggesting a robust understanding of query ordering and limits. Tapex has an average performance, especially in Projection, Comparison, and Selection tasks, though it struggles slightly with Operators, indicating a potential area for improvement in understanding complex operations. Tapas, while performing well in basic query components, significantly lags in Group By operations, highlighting a critical lack of expressiveness. SQuALL presents a balanced performance with a slight edge in Aggregation tasks but faces challenges in accurately interpreting Operator commands. 
%
Among our models, PCS exhibits the best overall performance (as on the test set), thanks to its robust handling of query types. Surprisingly, it however performs worse on group-by queries compared to models that include GB in $\mathcal{O}^*$. We suppose that this might be due to the variance due to the limited number of queries of that type. 
Finally, our model exhibits a pattern where simpler operators (projection, comparison, selection) are better handled when generated directly, while others (order by, aggregation, operators) do benefit from being executed externally. 
% Moreover, our models improve substantially the performance for queries containing operators with respect to OmniTab.

% -------- TABLE PER QUERY TYPE
\begin{table*}[t]

\centering
\caption{Performance (FDA) of models on the validation set, grouping results per type of query, for the models based on pre-order linearization (no alias). The column ALL reports FDA averaged over validation queries. Best results are in bold.}

% (no perturbation)
\resizebox{\textwidth}{!}{%
\begin{tabular}{lccccccccc}
\hline
{\small  \textbf{Model}} & {\small \textbf{Projection (ALL)}} & {\small \textbf{Comparison}} & {\small \textbf{Selection}} & {\small \textbf{Group By}} & {\small \textbf{Order By}} & {\small \textbf{Aggregation}} & {\small \textbf{Operator}} & {\small \textbf{Limit}} & {\small  $\sigma$} \\
\hline
\# & 500 & 367 & 363 & 30 & 151 & 206 & 75 & 153 & \\
\hline
Tapas  &  52.6 &  51.8 &  52.3 &  16.7 &  53.0 &  43.7 &  30.7 &  52.3 &  13.5 \\
Tapex   &  55.2 &  55.9 &  56.5 &  50.0 &  60.9 &  38.8 &  44.0 &  60.8 &   7.9 \\
Omnitab  &  \textbf{58.8} &  \textbf{59.7} &  \textbf{59.8} &  \textbf{56.7} &  {61.6} &  47.1 &  45.3 &  60.8 &   6.4 \\
\hline
P &  44.6 &  40.9 &  41.3 &  40.0 &  49.7 &  43.7 &  28.0 &  49.0 &   6.8 \\
+C          &  51.6 &  50.1 &  50.7 &  23.3 &  48.3 &  50.0 &  38.7 &  47.7 &   9.7 \\
+S         &  58.6 &  58.0 &  58.4 &  40.0 &  58.3 &  \textbf{52.4} &  \textbf{52.0} &  57.5 &   6.4 \\
+GB+H      &  57.8 &  57.8 &  58.4 &  23.3 &  57.0 &  49.5 &  49.3 &  56.2 &  11.8 \\
+OB    &  57.6 &  57.5 &  57.8 &  53.3 &  58.9 &  51.5 &  50.7 &  58.2 &   \underline{3.3} \\
+A   &  58.0 &  57.8 &  58.4 &  \textbf{56.7} &  \textbf{62.2} &  47.1 &  49.3 &  \textbf{61.4} &   5.4 \\
+OP &  56.6 &  57.8 &  58.4 &  50.0 &  60.3 &  46.1 &  42.7 &  60.1 &   6.8 \\

\hline
\end{tabular}%
}
\caption{Using validation data \label{tab:sensitivity:op} -- the row \# contains the number of matching queries (see Section \ref{sec:types})}

% (with perturbation)
\resizebox{\textwidth}{!}{%
\begin{tabular}{lccccccccc}
\hline
{\small  \textbf{Model}} &  {\small \textbf{Projection (ALL)}} & {\small \textbf{Comparison}} & {\small \textbf{Selection}} & {\small \textbf{Group By}} & {\small \textbf{Order By}} & {\small \textbf{Aggregation}} & {\small \textbf{Operator}} & {\small \textbf{Limit}} & {\small  $\sigma$} \\
\hline
Tapas  &  42.6 &  41.7 &  42.2 &  16.7 &  38.4 &  37.9 &  18.7 &  37.9 &  10.6 \\
Tapex       &  43.4 &  43.0 &  43.5 &  43.3 &  44.4 &  35.4 &  29.3 &  44.4 &   5.5 \\
Omnitab     &  45.4 &  44.7 &  44.9 &  36.7 &  42.4 &  39.3 &  30.7 &  42.5 &   5.1 \\
\hline
P   &  43.2 &  39.8 &  40.2 &  36.7 &  45.0 &  44.2 &  28.0 &  44.4 &   5.7 \\
+C   &  49.0 &  46.3 &  46.8 &  23.3 &  45.7 &  48.5 &  38.7 &  45.1 &   8.5 \\
+S         &  \textbf{53.6} &  \textbf{51.0} &  \textbf{51.2} &  40.0 &  \textbf{49.0} &  \textbf{51.9} &  \textbf{50.7} &  \textbf{48.4} &   4.2 \\
+GB+H      &  51.6 &  49.6 &  50.1 &  23.3 &  45.0 &  49.5 &  48.0 &  44.4 &   9.2 \\
+OB    &  50.6 &  50.7 &  51.2 &  40.0 &  43.7 &  48.1 &  46.7 &  43.1 &   4.1 \\
+A   &  47.2 &  46.0 &  46.6 &  \textbf{50.0} &  45.7 &  41.8 &  40.0 &  45.1 &   \underline{3.1} \\
+OP &  47.8 &  47.7 &  48.2 &  \textbf{50.0} &  45.7 &  43.2 &  30.7 &  45.8 &   6.1 \\

\hline
\end{tabular}%
}
\caption{Using validation data with \emph{random permutations} of each column \label{tab:sensitivity:op:perturbed} (see Section \ref{sec:perturbations})}

\end{table*}
% /-------- TABLE PER QUERY TYPE

Finally, table \ref{tab:sensibility:complexity} presents the performance of models with respect to the complexity of the query, as measured by the number of operators in the original computational graph. OmniTab and our models (especially +GB+H) demonstrate resilience with relatively stable performance across operation ranges. Tapex and Tapas, however, show a decline in performance as complexity increases, with Tapas notably struggling in the 8+ operation category showing the limit of extractive methods. 

\begin{table}[b]
\centering
\caption{Performance (FDA) with respect to the number of operators}
\label{tab:system_performance}
\begin{tabular}{lcccc}
\hline
\textbf{Model} & \textbf{1-4} & \textbf{4-8} & \textbf{8+}\\
\hline
Tapex       &  65.5 &  44.3 &  55.2 \\
Tapas       &  66.5 &  49.0 &  32.4 \\
Omnitab     &  65.0 &  \textbf{54.2} &  55.2 \\
\hline
+P           &  53.2 &  42.2 &  32.4 \\
+C          &  61.1 &  46.4 &  42.9 \\
+S         &  67.0 &  53.7 &  51.4 \\
+GB+H      &  \textbf{67.5} &  49.5 &  54.3 \\
+OB        &  63.6 &  52.6 &  55.2 \\
+A         &  65.0 &  53.1 &  \textbf{57.1} \\
+OP        &  63.1 &  50.0 &  56.2 \\

\hline
\end{tabular}
\label{tab:sensibility:complexity}
\end{table}

\subsection{Comparing linearization methods}

\begin{figure}[t]
  \centering
  \includegraphics[width=0.48\textwidth]{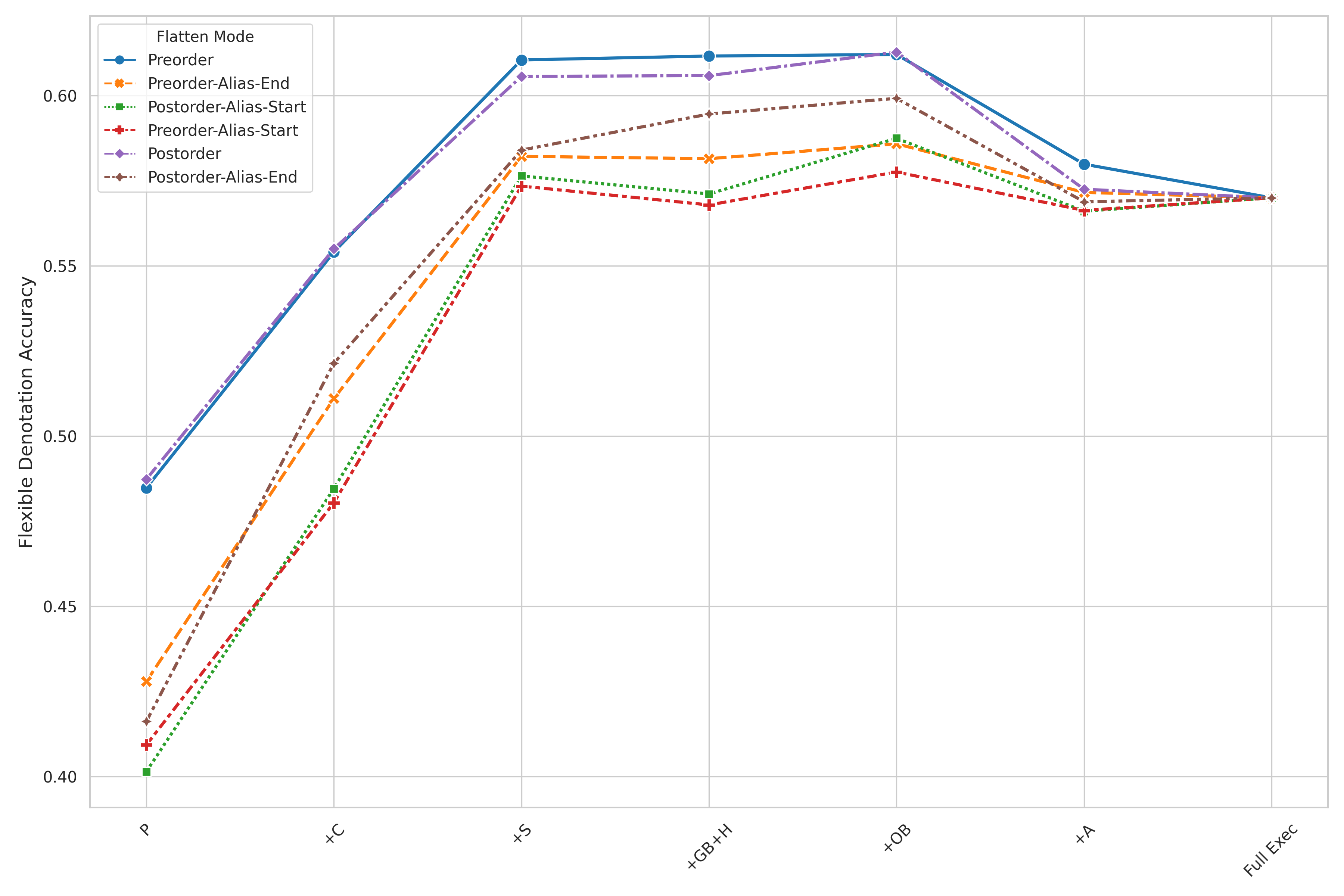}
  \caption{Evolution of FDA (test set) for different model variants. 
  % No-Exec corresponds to SQuALL.
  }
  \label{fig:linearization}
\end{figure}

In figure \ref{fig:linearization}, we show the impact of the linearization on the performance of the models. 
We can first observe that differences between our model variants decrease as most of the computational graph is executed, which was expected. 
Contrary to our expectations, however, using aliases has a negative impact, especially when they are more used (+P to +P+C+S), which shows that having too many aliases is problematic when generating a relational formula.
% -- this might be tackled using data augmentation techniques.
%
When using aliases, putting the tables after the operators did somehow improve the results. 
We think that these results might change with better training procedures (e.g. data augmentation with perturbations): we observed that models using aliases were more robust, but their overall performance was nevertheless below that of non-aliases ones.
Finally, we observe that there is a granularity level (+P+C+S) that achieves the best performance, corresponding to cases where only basic table selection is performed; moreover, this level is less prone to overfitting as discussed in Section \ref{sec:perturbations}.

% In exploring various strategies to optimize the graph's representation, we experimented with altering the traversal mode between preorder and postorder, as well as adjusting the aliasing approach. Additionally, we analyzed the impact of positioning operands either at the beginning or end of the sequence. Our findings suggest that placing operands at the sequence's end generally leads to slightly improved performance, though the difference compared to starting positions is not markedly significant. Regarding the aliasing method, which eliminates repetition by naming operators and operands, it appears to challenge the model's ability to discern a consistent logic. As for the comparison between preorder and postorder traversals, our observations indicate that both approaches yield comparable results in terms of performance. Globaly the transformers seems to prefer preorder flattenning methods over the others. 

\subsection{Ensembling}
\label{sec:ensembling}

\begin{figure}[t]
  \centering
  \includegraphics[width=0.48\textwidth]{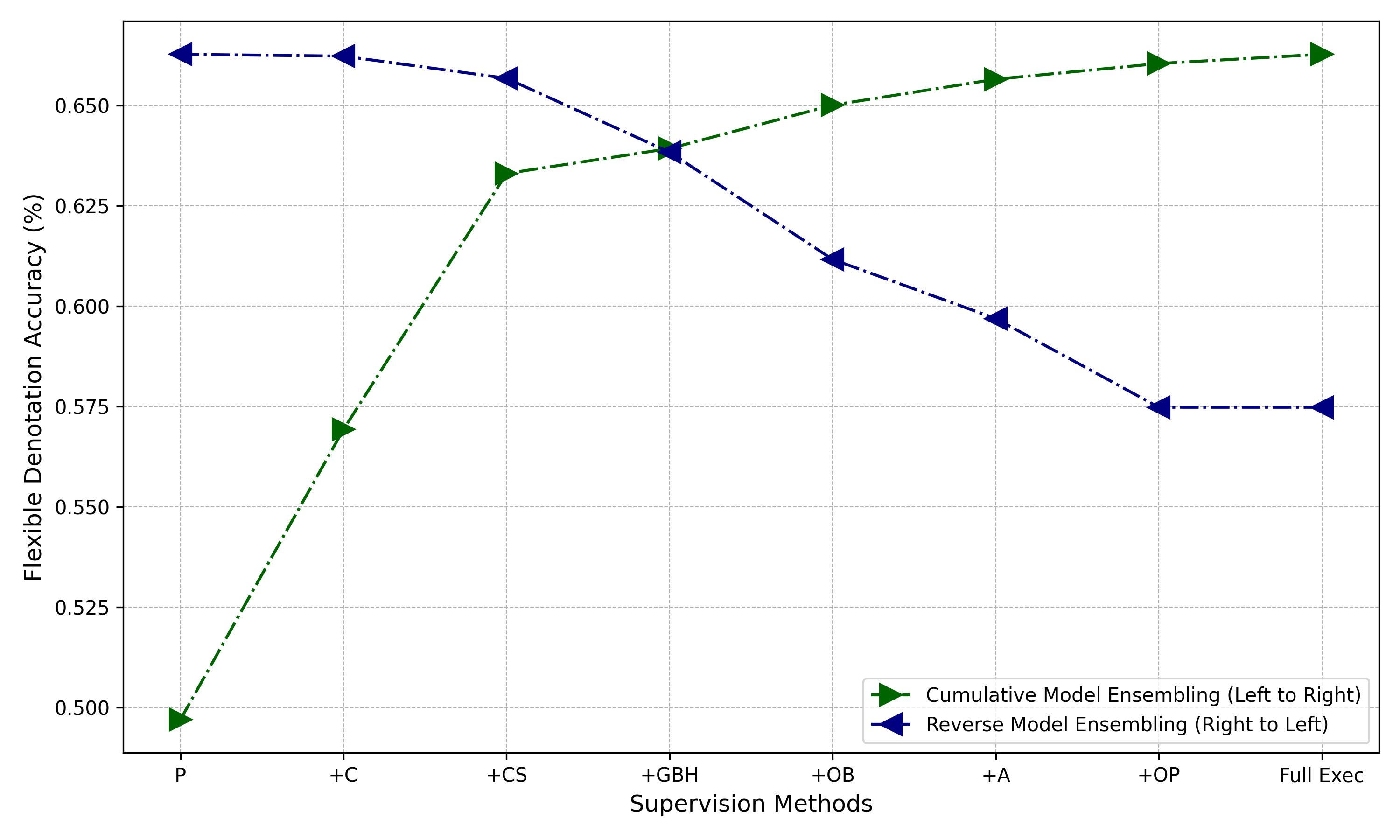}
  \caption{Evolution of FDA (test set) depending on the number of model variations in the ensemble. We either add models to the ensemble starting from the left (green) or the right (blue).}
  \label{fig:ensemble}
\end{figure}

Figure \ref{fig:ensemble} illustrates the results that we obtained using different ensembling combinations. The ensemble prediction is given by a majority vote. In case of ties, we use the validation FDA to weights the votes. 
We experimented with two ensembling settings: going from semantic parsing models to full execution, or in the opposite direction, i.e. from full execution to semantic parsing.  

First, performance improves whatever the ensembling method. This improvement can be explained with the analysis presented in Table \ref{tab:sensitivity:op:perturbed}, where we analyzed the performance depending on the operators composing the computational graph. 
%
% We highlight that these models have good performances over a large array of questions. 
While certain models excel in specific types of operations, others may show superiority in different areas. Such diversity among the models is important for ensembling.
% The ensembling method will mitigating the variance inherent between individual models, thereby signigicantly enhancing overall performance. 

%some are better for one type of operation and some for other. This diversity in these models allow the ensembling method to eliminate this variance between models and so to improve the performances.

%Specifically, each model has better performance for a different subset of queries. 

% Figure \ref{fig:ensemble} illustrates the evolution of model performance depending on the number of model variations in the ensemble (we use a vote weighted by model performance on the validation set).  Whatever the way of ensembling models, the performances increase -- except, for "+OP" and "Full", most probably because their behavior is very close. Altogether, those results show that our model variants are complementary, and suggest that more work is needed to understand what their relative strengths are and how to combine them into a single model.

\subsection{Sensitivity over table column cells perturbations}
\label{sec:perturbations}

Transformers architecture can easily overfit, especially in the case of a dataset like WTQ. 
% Even with our simple training procedure, we supposed that using algebra would make our model more robust.
%
To measure the importance of overfitting, we use the validation set (since the test set has no associated SQL queries) and perform random perturbations, i.e. we permute rows within each column. To avoid problems relative to the maximum length of the input, we ensure those perturbations only affect the parts present in the input of the transformer -- all models would have been affected, and this would have reduced the sensibility of our measures.

Results are shown in table \ref{tab:sensitivity:op:perturbed} using a pre-order (no alias) linearization (our best linearization method).
We observe that perturbation strongly affects even the best-performing approaches, as OmniTAB performance lowers from 58.8 to 45.4 (-13.4), Tapex from 55.2 to 43.4 (-11.8), and Tapas from 52.6 to 42.6 (-10.0). Our models are much less impacted. For instance, our best-performing approach (PCS) decreases its performance from 58.6 to 53.6 (-5.0), and beats the best baseline, Omnitab, by a large margin (53.6 vs 45.4), showing that data augmentation is less effective in preventing overfitting than generating formulas combining content and relational operators.

Among our models, we note that the lesser the amount of executed parts in the computation graph, the lower the decrease. As some models were initially more performant than others, we can note that the "P+C+S" model is the most effective one, with an average FDA of 53.6.
Finally, we can see that the impact on some operators (e.g. group by, limit, comparisons) is even higher for models where most or all of the computational graph is executed.

\section{Conclusion}

We explored the realm between semantic parsing and direct output generation for table QA, showing that PLMs can leverage an appropriate level of granularity where basic table manipulations (cleaning, selection) can be handled by the transformer itself while higher-level operations (e.g. aggregation, arithmetic) are better handled by dedicated tools. We showed that a model, appropriately trained, achieves a high performance compared to state-of-the-art, and that, more importantly, most PLMs baselines are prone to overfitting (by using a simple permutation of table cells), while our method is much less affected and beats the best baseline, OmniTab, by a wide margin. Future works will include more sophisticated training procedures, a sparse attention mechanism to cope with long tables such as LLMs, and more in-depth error analysis.

% --- Can be after the 8th page limit

\afterpage{\clearpage}
\section{Limitations \& Risks}

Our models have not been trained with data augmentation, which would help them to make them more robust -- even if other models could benefit from it (e.g. Tapex or Tapas), we hypothesize that it would have an even bigger impact on our model (The best baseline, OmniTab, was already trained with augmented data). Experimenting with more datasets would also have strengthened our results. However, as for all works on Table QA, WikiTableQuestion is still a resource of reference.

We did not compare thoroughly our results with LLMs but did report the results from the original papers. However, the gap between the best-performing LLMs and our model is not that high, showing the potential benefit of using partially executed formulas. Future works could include the fine-tuning of LLMs with our proposed supervision. 

Risks involved in this research are similar to those incoming from any NLP research, as an automatic understanding of data can be used maliciously, e.g. leaking confidential information from tables. However, this work focuses on an exploratory study of learning abilities, which is dedicated to the scientific community only.

\afterpage{\clearpage}
\section{Aknowledgements}
This work was partly funded by the ANR-21-CE23-0007 ACDC project. Experiments were performed using HPC resources from GENCI-IDRIS (Grant 2023-AD011014110).

\bibliographystyle{unsrtnat}
\bibliography{references}  %%% Uncomment this line and comment out the ``thebibliography'' section below to use the external .bib file (using bibtex) .

\begin{thebibliography}{35}
\providecommand{\natexlab}[1]{#1}
\providecommand{\url}[1]{\texttt{#1}}
\expandafter\ifx\csname urlstyle\endcsname\relax
  \providecommand{\doi}[1]{doi: #1}\else
  \providecommand{\doi}{doi: \begingroup \urlstyle{rm}\Url}\fi

\bibitem[Jin et~al.(2022)Jin, Siebert, Li, and Chen]{jin2022survey}
Nengzheng Jin, Joanna Siebert, Dongfang Li, and Qingcai Chen.
\newblock A survey on table question answering: recent advances.
\newblock In \emph{China Conference on Knowledge Graph and Semantic Computing}, pages 174--186. Springer, 2022.

\bibitem[Shi et~al.(2020)Shi, Zhao, Boyd-Graber, Daum{\'e}~III, and Lee]{shi2020potential}
Tianze Shi, Chen Zhao, Jordan Boyd-Graber, Hal Daum{\'e}~III, and Lillian Lee.
\newblock On the potential of lexico-logical alignments for semantic parsing to sql queries.
\newblock \emph{arXiv preprint arXiv:2010.11246}, 2020.

\bibitem[Pasupat and Liang(2015)]{pasupat2015compositional}
Panupong Pasupat and Percy Liang.
\newblock Compositional semantic parsing on semi-structured tables.
\newblock \emph{arXiv preprint arXiv:1508.00305}, 2015.

\bibitem[Liu et~al.(2021)Liu, Chen, Guo, Ziyadi, Lin, Chen, and Lou]{liu2021tapex}
Qian Liu, Bei Chen, Jiaqi Guo, Morteza Ziyadi, Zeqi Lin, Weizhu Chen, and Jian-Guang Lou.
\newblock Tapex: Table pre-training via learning a neural sql executor.
\newblock \emph{arXiv preprint arXiv:2107.07653}, 2021.

\bibitem[Herzig et~al.(2020)Herzig, Nowak, M{\"u}ller, Piccinno, and Eisenschlos]{herzig2020tapas}
Jonathan Herzig, Pawe{\l}~Krzysztof Nowak, Thomas M{\"u}ller, Francesco Piccinno, and Julian~Martin Eisenschlos.
\newblock Tapas: Weakly supervised table parsing via pre-training.
\newblock \emph{arXiv preprint arXiv:2004.02349}, 2020.

\bibitem[Zhou et~al.(2022{\natexlab{a}})Zhou, Bao, Duan, Wu, He, and Zhao]{zhou2022unirpg}
Yongwei Zhou, Junwei Bao, Chaoqun Duan, Youzheng Wu, Xiaodong He, and Tiejun Zhao.
\newblock Unirpg: Unified discrete reasoning over table and text as program generation.
\newblock \emph{arXiv preprint arXiv:2210.08249}, 2022{\natexlab{a}}.

\bibitem[Pruksachatkun et~al.(2020)Pruksachatkun, Phang, Liu, Htut, Zhang, Pang, Vania, Kann, and Bowman]{pruksachatkun2020intermediate}
Yada Pruksachatkun, Jason Phang, Haokun Liu, Phu~Mon Htut, Xiaoyi Zhang, Richard~Yuanzhe Pang, Clara Vania, Katharina Kann, and Samuel~R Bowman.
\newblock Intermediate-task transfer learning with pretrained models for natural language understanding: When and why does it work?
\newblock \emph{arXiv preprint arXiv:2005.00628}, 2020.

\bibitem[Geva et~al.(2020)Geva, Gupta, and Berant]{geva2020injecting}
Mor Geva, Ankit Gupta, and Jonathan Berant.
\newblock Injecting numerical reasoning skills into language models.
\newblock \emph{arXiv preprint arXiv:2004.04487}, 2020.

\bibitem[Yu et~al.(2020)Yu, Wu, Lin, Wang, Tan, Yang, Radev, Socher, and Xiong]{yu2020grappa}
Tao Yu, Chien-Sheng Wu, Xi~Victoria Lin, Bailin Wang, Yi~Chern Tan, Xinyi Yang, Dragomir Radev, Richard Socher, and Caiming Xiong.
\newblock Grappa: Grammar-augmented pre-training for table semantic parsing.
\newblock \emph{arXiv preprint arXiv:2009.13845}, 2020.

\bibitem[Eisenschlos et~al.(2021)Eisenschlos, Gor, M{\"u}ller, and Cohen]{eisenschlos2021mate}
Julian~Martin Eisenschlos, Maharshi Gor, Thomas M{\"u}ller, and William~W Cohen.
\newblock Mate: multi-view attention for table transformer efficiency.
\newblock \emph{arXiv preprint arXiv:2109.04312}, 2021.

\bibitem[Golchin and Surdeanu(2023)]{golchin2023time}
Shahriar Golchin and Mihai Surdeanu.
\newblock Time travel in llms: Tracing data contamination in large language models.
\newblock \emph{arXiv preprint arXiv:2308.08493}, 2023.

\bibitem[Wang et~al.(2021)Wang, Dong, Jia, Li, Fu, Han, and Zhang]{wang2021tuta}
Zhiruo Wang, Haoyu Dong, Ran Jia, Jia Li, Zhiyi Fu, Shi Han, and Dongmei Zhang.
\newblock Tuta: Tree-based transformers for generally structured table pre-training.
\newblock In \emph{Proceedings of the 27th ACM SIGKDD Conference on Knowledge Discovery \& Data Mining}, pages 1780--1790, 2021.

\bibitem[Chen(2022)]{chen2022large}
Wenhu Chen.
\newblock Large language models are few (1)-shot table reasoners.
\newblock \emph{arXiv preprint arXiv:2210.06710}, 2022.

\bibitem[Cheng et~al.(2022)Cheng, Xie, Shi, Li, Nadkarni, Hu, Xiong, Radev, Ostendorf, Zettlemoyer, et~al.]{cheng2022binding}
Zhoujun Cheng, Tianbao Xie, Peng Shi, Chengzu Li, Rahul Nadkarni, Yushi Hu, Caiming Xiong, Dragomir Radev, Mari Ostendorf, Luke Zettlemoyer, et~al.
\newblock Binding language models in symbolic languages.
\newblock \emph{arXiv preprint arXiv:2210.02875}, 2022.

\bibitem[Wang et~al.(2024)Wang, Zhang, Li, Eisenschlos, Perot, Wang, Miculicich, Fujii, Shang, Lee, et~al.]{wang2024chain}
Zilong Wang, Hao Zhang, Chun-Liang Li, Julian~Martin Eisenschlos, Vincent Perot, Zifeng Wang, Lesly Miculicich, Yasuhisa Fujii, Jingbo Shang, Chen-Yu Lee, et~al.
\newblock Chain-of-table: Evolving tables in the reasoning chain for table understanding.
\newblock \emph{arXiv preprint arXiv:2401.04398}, 2024.

\bibitem[Hu et~al.(2021)Hu, Shen, Wallis, Allen-Zhu, Li, Wang, Wang, and Chen]{hu2021lora}
Edward~J Hu, Yelong Shen, Phillip Wallis, Zeyuan Allen-Zhu, Yuanzhi Li, Shean Wang, Lu~Wang, and Weizhu Chen.
\newblock Lora: Low-rank adaptation of large language models.
\newblock \emph{arXiv preprint arXiv:2106.09685}, 2021.

\bibitem[Dettmers et~al.(2024)Dettmers, Pagnoni, Holtzman, and Zettlemoyer]{dettmers2024qlora}
Tim Dettmers, Artidoro Pagnoni, Ari Holtzman, and Luke Zettlemoyer.
\newblock Qlora: Efficient finetuning of quantized llms.
\newblock \emph{Advances in Neural Information Processing Systems}, 36, 2024.

\bibitem[Jiang et~al.(2022)Jiang, Mao, He, Neubig, and Chen]{jiang2022omnitab}
Zhengbao Jiang, Yi~Mao, Pengcheng He, Graham Neubig, and Weizhu Chen.
\newblock Omnitab: Pretraining with natural and synthetic data for few-shot table-based question answering.
\newblock \emph{arXiv preprint arXiv:2207.03637}, 2022.

\bibitem[Schick et~al.(2024)Schick, Dwivedi-Yu, Dess{\`\i}, Raileanu, Lomeli, Hambro, Zettlemoyer, Cancedda, and Scialom]{schick2024toolformer}
Timo Schick, Jane Dwivedi-Yu, Roberto Dess{\`\i}, Roberta Raileanu, Maria Lomeli, Eric Hambro, Luke Zettlemoyer, Nicola Cancedda, and Thomas Scialom.
\newblock Toolformer: Language models can teach themselves to use tools.
\newblock \emph{Advances in Neural Information Processing Systems}, 36, 2024.

\bibitem[Wang et~al.(2019)Wang, Shin, Liu, Polozov, and Richardson]{wang2019rat}
Bailin Wang, Richard Shin, Xiaodong Liu, Oleksandr Polozov, and Matthew Richardson.
\newblock Rat-sql: Relation-aware schema encoding and linking for text-to-sql parsers.
\newblock \emph{arXiv preprint arXiv:1911.04942}, 2019.

\bibitem[Min et~al.(2019)Min, Chen, Hajishirzi, and Zettlemoyer]{min2019discrete}
Sewon Min, Danqi Chen, Hannaneh Hajishirzi, and Luke Zettlemoyer.
\newblock A discrete hard em approach for weakly supervised question answering.
\newblock \emph{arXiv preprint arXiv:1909.04849}, 2019.

\bibitem[Zhong et~al.(2017)Zhong, Xiong, and Socher]{zhong2017seq2sql}
Victor Zhong, Caiming Xiong, and Richard Socher.
\newblock Seq2sql: Generating structured queries from natural language using reinforcement learning.
\newblock \emph{arXiv preprint arXiv:1709.00103}, 2017.

\bibitem[Mueller et~al.(2019)Mueller, Piccinno, Nicosia, Shaw, and Altun]{mueller2019answering}
Thomas Mueller, Francesco Piccinno, Massimo Nicosia, Peter Shaw, and Yasemin Altun.
\newblock Answering conversational questions on structured data without logical forms.
\newblock \emph{arXiv preprint arXiv:1908.11787}, 2019.

\bibitem[Eisenschlos et~al.(2020)Eisenschlos, Krichene, and M{\"u}ller]{eisenschlos2020understanding}
Julian~Martin Eisenschlos, Syrine Krichene, and Thomas M{\"u}ller.
\newblock Understanding tables with intermediate pre-training.
\newblock \emph{arXiv preprint arXiv:2010.00571}, 2020.

\bibitem[Cheng et~al.(2021)Cheng, Dong, Jia, Wu, Han, Cheng, and Zhang]{cheng2021fortap}
Zhoujun Cheng, Haoyu Dong, Ran Jia, Pengfei Wu, Shi Han, Fan Cheng, and Dongmei Zhang.
\newblock Fortap: Using formulas for numerical-reasoning-aware table pretraining.
\newblock \emph{arXiv preprint arXiv:2109.07323}, 2021.

\bibitem[Zhou et~al.(2022{\natexlab{b}})Zhou, Hu, Dong, Cheng, Han, and Zhang]{zhou2022tacube}
Fan Zhou, Mengkang Hu, Haoyu Dong, Zhoujun Cheng, Shi Han, and Dongmei Zhang.
\newblock Tacube: Pre-computing data cubes for answering numerical-reasoning questions over tabular data.
\newblock \emph{arXiv preprint arXiv:2205.12682}, 2022{\natexlab{b}}.

\bibitem[Zhu et~al.(2021)Zhu, Lei, Huang, Wang, Zhang, Lv, Feng, and Chua]{zhu2021tat}
Fengbin Zhu, Wenqiang Lei, Youcheng Huang, Chao Wang, Shuo Zhang, Jiancheng Lv, Fuli Feng, and Tat-Seng Chua.
\newblock Tat-qa: A question answering benchmark on a hybrid of tabular and textual content in finance.
\newblock \emph{arXiv preprint arXiv:2105.07624}, 2021.

\bibitem[Codd(1970)]{codd1970relational}
Edgar~F Codd.
\newblock A relational model of data for large shared data banks.
\newblock \emph{Communications of the ACM}, 13\penalty0 (6):\penalty0 377--387, 1970.

\bibitem[Lewis et~al.(2019)Lewis, Liu, Goyal, Ghazvininejad, Mohamed, Levy, Stoyanov, and Zettlemoyer]{lewis2019bart}
Mike Lewis, Yinhan Liu, Naman Goyal, Marjan Ghazvininejad, Abdelrahman Mohamed, Omer Levy, Ves Stoyanov, and Luke Zettlemoyer.
\newblock Bart: Denoising sequence-to-sequence pre-training for natural language generation, translation, and comprehension.
\newblock \emph{arXiv preprint arXiv:1910.13461}, 2019.

\bibitem[Wei et~al.(2022)Wei, Wang, Schuurmans, Bosma, Xia, Chi, Le, Zhou, et~al.]{wei2022chain}
Jason Wei, Xuezhi Wang, Dale Schuurmans, Maarten Bosma, Fei Xia, Ed~Chi, Quoc~V Le, Denny Zhou, et~al.
\newblock Chain-of-thought prompting elicits reasoning in large language models.
\newblock \emph{Advances in Neural Information Processing Systems}, 35:\penalty0 24824--24837, 2022.

\bibitem[Yin et~al.(2020)Yin, Neubig, Yih, and Riedel]{yin2020tabert}
Pengcheng Yin, Graham Neubig, Wen-tau Yih, and Sebastian Riedel.
\newblock Tabert: Pretraining for joint understanding of textual and tabular data.
\newblock \emph{arXiv preprint arXiv:2005.08314}, 2020.

\bibitem[Yang et~al.(2022)Yang, Gupta, Upadhyay, He, Goel, and Paul]{yang2022tableformer}
Jingfeng Yang, Aditya Gupta, Shyam Upadhyay, Luheng He, Rahul Goel, and Shachi Paul.
\newblock Tableformer: Robust transformer modeling for table-text encoding.
\newblock \emph{arXiv preprint arXiv:2203.00274}, 2022.

\bibitem[Krichene et~al.(2021)Krichene, M{\"u}ller, and Eisenschlos]{krichene2021dot}
Syrine Krichene, Thomas M{\"u}ller, and Julian~Martin Eisenschlos.
\newblock Dot: An efficient double transformer for nlp tasks with tables.
\newblock \emph{arXiv preprint arXiv:2106.00479}, 2021.

\bibitem[Zhao et~al.(2022)Zhao, Nan, Qi, Zhang, and Radev]{zhao2022reastap}
Yilun Zhao, Linyong Nan, Zhenting Qi, Rui Zhang, and Dragomir Radev.
\newblock Reastap: Injecting table reasoning skills during pre-training via synthetic reasoning examples.
\newblock \emph{arXiv preprint arXiv:2210.12374}, 2022.

\bibitem[Ye et~al.(2023)Ye, Hui, Yang, Li, Huang, and Li]{ye2023large}
Yunhu Ye, Binyuan Hui, Min Yang, Binhua Li, Fei Huang, and Yongbin Li.
\newblock Large language models are versatile decomposers: Decompose evidence and questions for table-based reasoning.
\newblock \emph{arXiv preprint arXiv:2301.13808}, 2023.

\end{thebibliography}

\appendix

\section{Appendix}
\label{sec:appendix}

\subsection{Linearization}
\label{sec:appendix:linearization}

\subsubsection{Using aliases} \label{sec:appendix:aliases}

In the case of pre-order, we denote the alias for node $n$ with $\alpha_n$ and use the following linearization:
\begin{multline}
l_{pre}(n)= 
\begin{cases}
l(x_n) \oplus_i \alpha_{n_i} \oplus \alpha_n \oplus_i l^c_{pre}(n_i) \\ \qquad \mbox{if}\ x \in \mathcal{O} \\
\varnothing \qquad\text{else}
\end{cases}
\end{multline}
where $\varnothing$ denotes the empty sequence and $l^c$ is either empty -- if the operator has already been linearized -- or $|| \oplus l_{pre}(n_i)$ if not. Note that the order of linearization is important, but to avoid more complicated notations we do not make it explicit here.

\label{app:linearization}

\begin{table*}
\centering
  \begin{tabular}{ m{2cm} m{8cm} m{5cm}}
    \hline
    $\mathcal{O}^*$ & Logical Form & Graph \\
    \hline
    \{P\} & \textcolor{red}{Limit 1} || \textcolor{red}{OB desc} || \textcolor{red}{GB} || \textcolor{red}{WHERE} || \textcolor{blue}{newtongrange | .. | fauldhouse} || \textcolor{red}{IN 'fauldhouse', 'newtongrange'} || \textcolor{blue}{newtongrange | .. | fauldhouse} || \textcolor{red}{WHERE} || \textcolor{blue}{newtongrange| .. | fauldhouse} || \textcolor{red}{IN 'fauldhouse', 'newtongrange'} || \textcolor{blue}{newtongrange| .. | fauldhouse} || \textcolor{red}{COUNT} || \textcolor{red}{GB} || \textcolor{red}{WHERE} || \textcolor{blue}{newtongrange| .. | fauldhouse} || \textcolor{red}{in 'fauldhouse', 'newtongrange'} || \textcolor{blue}{newtongrange| .. | fauldhouse} || \textcolor{red}{WHERE} || \textcolor{blue}{1 | .. | 7} || \textcolor{red}{IN 'fauldhouse', 'newtongrange'} || \textcolor{blue}{newtongrange | .. | fauldhouse} || & \includegraphics[width=5cm]{Omega.png}  \\
    \hline
    \{P, C\} & \textcolor{red}{Limit 1} || \textcolor{red}{OB desc} || \textcolor{red}{GB} || \textcolor{red}{WHERE} || \textcolor{blue}{newtongrange | .. | fauldhouse} || \textcolor{blue}{t | f | .. | t | t} || \textcolor{red}{WHERE} || \textcolor{blue}{newtongrange| .. | fauldhouse} || \textcolor{blue}{t | f | .. | t | t} || \textcolor{red}{COUNT} || \textcolor{red}{GB} || \textcolor{red}{WHERE} || \textcolor{blue}{newtongrange| .. | fauldhouse} || \textcolor{blue}{t | f | .. | t | t} || \textcolor{red}{WHERE} || \textcolor{blue}{1 | .. | 7} || \textcolor{blue}{t | f | .. | t | t} || & \includegraphics[width=5cm]{OMEGAPC.png}\\
    \hline
    \{P, C, S\} & \textcolor{red}{Limit 1} || \textcolor{red}{OB desc} || \textcolor{red}{GB} || \textcolor{blue}{newtongrange | fauldhouse | fauldhouse}  || \textcolor{blue}{newtongrange | fauldhouse | fauldhouse}  || \textcolor{red}{COUNT} || \textcolor{red}{GB}  || \textcolor{blue}{newtongrange | fauldhouse | fauldhouse}  || \textcolor{blue}{1 | 2 | 7} || & \includegraphics[height=1.7cm,width=3.8cm]{OMEGAPCS.png}\\
    \hline
    \{P, C, S, GB, H\} & \textcolor{red}{Limit 1} || \textcolor{red}{OB desc} || \textcolor{blue}{fauldhouse,, fauldhouse | newtongrange} || \textcolor{red}{COUNT} || \textcolor{blue}{2,, 7 | 1} || & \includegraphics[height=1.6cm,width=2.35cm]{OMEGAPCSGBH.png}\\
    \hline
    \{P, C, S, GB, H, OB\} & \textcolor{red}{Limit 1} || \textcolor{red}{OB desc} || \textcolor{blue}{fauldhouse,, fauldhouse | newtongrange} || \textcolor{red}{COUNT} || \textcolor{blue}{2,, 7 | 1} || & \includegraphics[height=1.6cm,width=2.35cm]{OMEGAPCSGBH.png}\\
    \hline
    \{P, C, S, GB, H, OB, A\} & \textcolor{red}{Limit 1} || \textcolor{blue}{fauldhouse,, fauldhouse | newtongrange} || & \includegraphics[height=0.65cm,width=1.5cm]{OMEGAPCSGBHOBA.png}\\
    \hline
    % \{P, C, S, GB, H, OB, A, OP, L\} & \textcolor{blue}{fauldhouse}  & \includegraphics[height=0.65cm,width=0.63cm]{latex/Image/OMEGAPCSGBHOBAL.drawio.png}\\
    Full & \textcolor{blue}{fauldhouse}  &  \\

    \hline
\end{tabular}
\caption{\label{tab:example:preorder}
Example of Pre-order linearization for the query  "SELECT East Region FROM w WHERE East Region in ('fauldhouse', 'newtongrange') GROUP BY East Region ORDER BY COUNT ( id ) DESC LIMIT 1" Natural Language question ="'which team has made the roll of honour more times in the east region south division: fauldhouse or newtongrange ?'"
}
\end{table*}

\begin{table*}
\centering
  \begin{tabular}{ m{2cm} m{8cm} m{5cm}}
    \hline
    $\mathcal{O}^*$ & Logical Form & Graph \\

    \hline
    \{P\} &  \textcolor{blue}{newtongrange | .. | fauldhouse} || \textcolor{red}{IN 'fauldhouse', 'newtongrange'} || \textcolor{blue}{1 | .. | 7} || \textcolor{red}{WHERE} || \textcolor{blue}{newtongrange| .. | fauldhouse} || \textcolor{red}{in 'fauldhouse', 'newtongrange'} || \textcolor{blue}{newtongrange| .. | fauldhouse} || \textcolor{red}{WHERE} || \textcolor{red}{GB} || \textcolor{red}{COUNT} || \textcolor{blue}{newtongrange| .. | fauldhouse} || \textcolor{red}{IN 'fauldhouse', 'newtongrange'} || \textcolor{blue}{newtongrange| .. | fauldhouse} || \textcolor{red}{WHERE} || \textcolor{blue}{newtongrange | .. | fauldhouse} || \textcolor{red}{IN 'fauldhouse', 'newtongrange'} || \textcolor{blue}{newtongrange | .. | fauldhouse} || \textcolor{red}{WHERE} || \textcolor{red}{GB} || \textcolor{red}{OB desc} || \textcolor{red}{Limit 1} || & \includegraphics[width=5cm]{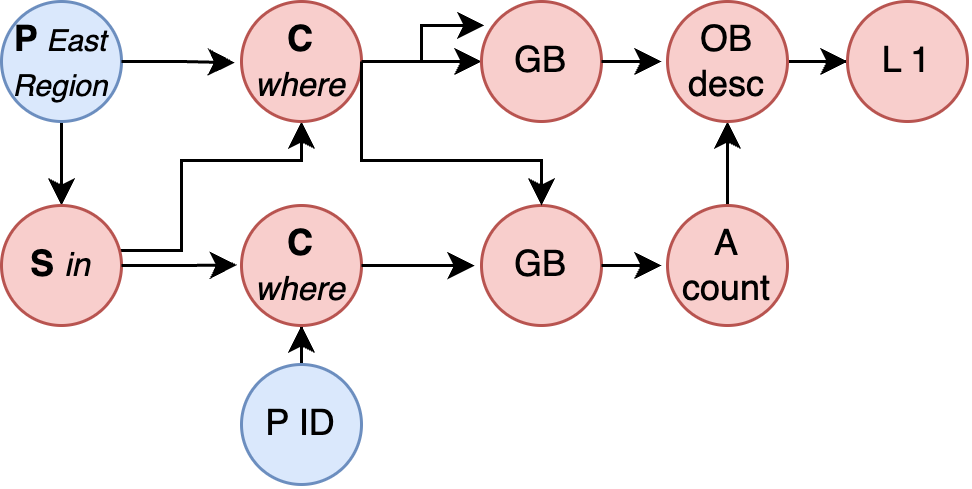}  \\
    \hline

    \{P, C\} &  \textcolor{blue}{t | f | .. | t | t} || \textcolor{blue}{1 | .. | 7} || \textcolor{red}{WHERE} || \textcolor{blue}{t | f | .. | t | t} || \textcolor{blue}{newtongrange| .. | fauldhouse} || \textcolor{red}{WHERE} || \textcolor{red}{GB} || \textcolor{red}{COUNT} || \textcolor{blue}{t | f | .. | t | t} || \textcolor{blue}{newtongrange| .. | fauldhouse} || \textcolor{red}{WHERE} || \textcolor{blue}{t | f | .. | t | t} || \textcolor{blue}{newtongrange | .. | fauldhouse} || \textcolor{red}{WHERE} || \textcolor{red}{GB} || \textcolor{red}{OB desc} || \textcolor{red}{Limit 1} ||& \includegraphics[width=5cm]{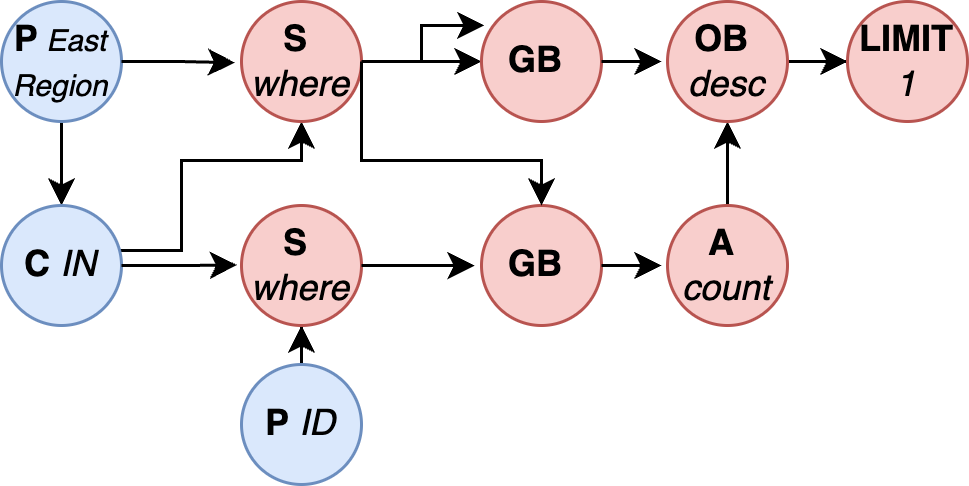}\\
    \hline
    \{P, C, S\} & \textcolor{blue}{1 | 2 | 7} || \textcolor{blue}{newtongrange | fauldhouse | fauldhouse}  || \textcolor{red}{GB}  || \textcolor{red}{COUNT} || \textcolor{blue}{newtongrange | fauldhouse | fauldhouse}  || \textcolor{blue}{newtongrange | fauldhouse | fauldhouse}  || \textcolor{red}{GB} || \textcolor{red}{OB desc} || \textcolor{red}{Limit 1} || & \includegraphics[height=1.7cm,width=3.8cm]{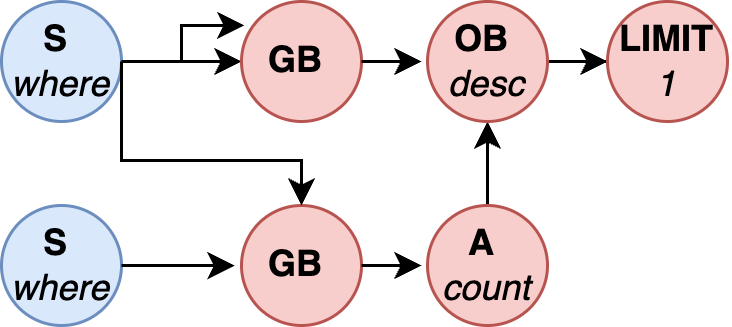}\\
    \hline
    \{P, C, S, GB, H\} &  \textcolor{blue}{2,, 7 | 1} || \textcolor{red}{COUNT} || \textcolor{blue}{fauldhouse,, fauldhouse | newtongrange} || \textcolor{red}{OB desc} ||\textcolor{red}{Limit 1} || & \includegraphics[height=1.6cm,width=2.35cm]{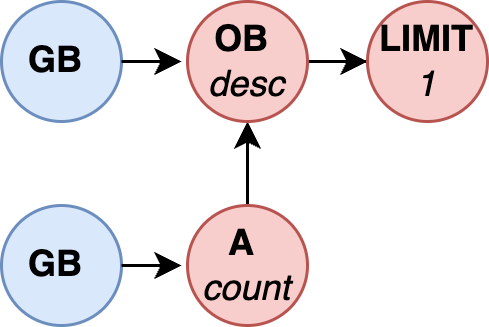}\\
    \hline
    \{P, C, S, GB, H, OB\} & \textcolor{blue}{2,, 7 | 1} || \textcolor{red}{COUNT} || \textcolor{blue}{fauldhouse,, fauldhouse | newtongrange} || \textcolor{red}{OB desc} ||\textcolor{red}{Limit 1} || & \includegraphics[height=1.6cm,width=2.35cm]{omegapcsgbh.png}\\
    \hline
    \{P, C, S, GB, H, OB, A\} & \textcolor{blue}{fauldhouse,, fauldhouse | newtongrange} || \textcolor{red}{Limit 1} || & \includegraphics[height=0.65cm,width=1.5cm]{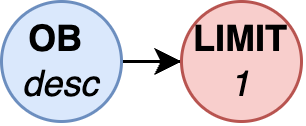}\\
    \hline

    Full & \textcolor{blue}{fauldhouse}  &  \\
    \hline
\end{tabular}
\caption{\label{tab:example:postorder}
Example of Post-order linearization for the query : "SELECT East Region FROM w WHERE East Region in ('fauldhouse', 'newtongrange') GROUP BY East Region ORDER BY COUNT ( id ) DESC LIMIT 1". Natural Language question ="'which team has made the roll of honour more times in the east region south division: fauldhouse or newtongrange ?'"
}
\end{table*}

\begin{table*}
\centering
  \begin{tabular}{ m{2cm} m{8cm} m{5cm}}
    \hline
    $\mathcal{O}^*$ & Logical Form & Graph \\

    \hline
    \{P\} & \textcolor{red}{ N10 Limit 1 N8} || \textcolor{red}{N8 ob desc N12 N22} || \textcolor{red}{N12 gb N7 N16} || \textcolor{red}{N7 WHERE N24 N6} || \textcolor{red}{N24 in 'fauldhouse united', 'newtongrange star' N6} || \textcolor{red}{N16 WHERE N24 N6} || \textcolor{red}{N22 COUNT N34} || \textcolor{red}{N34 GB N7 N26} || \textcolor{red}{N26 WHERE N24 N33} ||| \textcolor{blue}{N6 newtongrange star| ... | fauldhouse united }|| \textcolor{blue}{N33 1| 2| ... | 7} || & \includegraphics[width=5cm]{Omega.png}  \\
    \hline

    \{P, C\} & \textcolor{red}{ N22 Limit 1 N10} || \textcolor{red}{N10 OB desc N14 N13 || N14 GB N1 N2} || \textcolor{red}{N1 WHERE N8 N32} || \textcolor{red}{N2 WHERE N8 N32} || \textcolor{red}{N13 count N11} || \textcolor{red}{N11 GB N1 N12} || \textcolor{red}{N12 WHERE N8 N15} ||| \textcolor{blue}{N32 newtongrange star| fauldhouse united| ..| dalkeith thistle| fauldhouse united} || \textcolor{blue}{N8 t| .. | t} || \textcolor{blue}{N15 1 | .. | 6| } ||& \includegraphics[width=5cm]{omegapc.png}\\
    \hline
    \{P, C, S\} & \textcolor{red}{N11 Limit 1 N31} || \textcolor{red}{N31 OB desc N29 N3} || \textcolor{red}{N29 GB N4 N4} || \textcolor{red}{N3 COUNT N6} || \textcolor{red}{N6 GB N4 N5} ||| \textcolor{blue}{N4 newtongrange star| fauldhouse united| fauldhouse united} || \textcolor{blue}{N5 1| 2| 7} || & \includegraphics[height=1.7cm,width=3.8cm]{omegapcs.png}\\
    \hline
    \{P, C, S, GB, H\} & \textcolor{red}{ N35 Limit 1 N38 }|| \textcolor{red}{N38 OB desc N29 N27} || \textcolor{red}{N27 COUNT N3} ||| \textcolor{blue}{N29 fauldhouse united,, fauldhouse united| newtongrange sta} || \textcolor{blue}{N3 2,, 7| 1 } || & \includegraphics[height=1.6cm,width=2.35cm]{omegapcsgbh.png}\\
    \hline
    \{P, C, S, GB, H, OB\} &  \textcolor{red}{N22 Limit 1 N32} ||  \textcolor{red}{N32 OB desc N36 N23} ||  \textcolor{red}{N23 COUNT N38} |||  \textcolor{blue}{N36 fauldhouse united,, fauldhouse united| newtongrange star} ||  \textcolor{blue}{N38 2,, 7| 1} || & \includegraphics[height=1.6cm,width=2.35cm]{omegapcsgbh.png}\\
    \hline
    \{P, C, S, GB, H, OB, A\} & \textcolor{red}{N37 Limit 1 N36} ||| \textcolor{blue}{N36 fauldhouse united,, fauldhouse united| newtongrange star} || & \includegraphics[height=0.65cm,width=1.5cm]{omegapcsgbhoba.png}\\
    \hline
    Full & \textcolor{blue}{fauldhouse}  & \\
    \hline
\end{tabular}
\caption{\label{tab:example:preorder-alias}
Example of pre-order with alias (tables at the end) linearization for query : "SELECT East Region FROM w WHERE East Region in ('fauldhouse', 'newtongrange') GROUP BY East Region ORDER BY COUNT ( id ) DESC LIMIT 1" Natural Language question ="'which team has made the roll of honour more times in the east region south division: fauldhouse or newtongrange ?'"
}
\end{table*}

\begin{table*}
\centering
  \begin{tabular}{ m{2cm} m{8cm} m{5cm}}
    \hline
    $\mathcal{O}^*$ & Logical Form & Graph \\

    \hline
    \{P\} &  \textcolor{blue}{N24 newtongrange star| ..| fauldhouse united} || \textcolor{blue}{N1 1| 2| ..| 7 }||| \textcolor{red}{N3 Limit 1 N17} || \textcolor{red}{N17 OB desc N10 N5 }|| \textcolor{blue}{N10 GB N11 N35} || \textcolor{red}{N11 WHERE N4 N24} || \textcolor{red}{N4 in 'fauldhouse united', 'newtongrange star' N24} || \textcolor{red}{N35 WHERE N4 N24} || \textcolor{red}{N5 COUNT N7} || \textcolor{red}{N7 GB N11 N12 }|| \textcolor{red}{N12 WHERE N4 N1} || & \includegraphics[width=5cm]{omega.png}  \\
    \hline

    \{P, C\} &\textcolor{blue}{N12 newtongrange star| .. | fauldhouse united} || \textcolor{blue}{N16 t| t| .. | f| t } || \textcolor{blue}{N27 1| .. | 7} ||| \textcolor{red}{N28 Limit 1 N38} || \textcolor{red}{N38 OB desc N5 N24} || \textcolor{red}{N5 GB N13 N18} || \textcolor{red}{N13 WHERE N16 N12} || \textcolor{red}{N18 WHERE N16 N12} || \textcolor{red}{N24 COUNT N7} || \textcolor{red}{N7 GB N13 N8} || \textcolor{red}{N8 WHERE N16 N27} ||& \includegraphics[width=5cm]{omegapc.png}\\
    \hline
    \{P, C, S\} &  \textcolor{blue}{N22 newtongrange star| .. | fauldhouse united} || \textcolor{blue}{N10 1| .. | 7} ||| \textcolor{red}{N16 Limit 1 N4} || \textcolor{red}{N4 OB desc N2 N29} || \textcolor{red}{N2 GB N22 N22} || \textcolor{red}{N29 COUNT N30} || \textcolor{red}{N30 GB N22 N10} || & \includegraphics[height=1.7cm,width=3.8cm]{omegapcs.png}\\
    \hline
    \{P, C, S, GB, H\} &  \textcolor{blue}{N17 fauldhouse united,, fauldhouse united| newtongrange star} || \textcolor{blue}{N10 2,, 7| 1} ||| \textcolor{red}{N37 Limit 1 N12} || \textcolor{red}{N12 OB desc N17 N21} || \textcolor{red}{N21 COUNT N10} || & \includegraphics[height=1.6cm,width=2.35cm]{omegapcsgbh.png}\\
    \hline
    \{P, C, S, GB, H, OB\} & \textcolor{blue}{N25 fauldhouse united,, fauldhouse united| newtongrange star} || \textcolor{blue}{N22 2,, 7| 1} ||| \textcolor{red}{N2 Limit 1 N8} ||\textcolor{red}{N8 OB desc N25 N24} || \textcolor{red}{N24 COUNT N22} || & \includegraphics[height=1.6cm,width=2.35cm]{omegapcsgbh.png}\\
    \hline
    \{P, C, S, GB, H, OB, A\} & \textcolor{blue}{N12 fauldhouse united,, fauldhouse united| newtongrange star} |||\textcolor{red}{N10 Limit 1 N12} || & \includegraphics[height=0.65cm,width=1.5cm]{omegapcsgbhoba.png}\\
    \hline
    % \{P, C, S, GB, H, OB, A, OP, L\} & \textcolor{blue}{fauldhouse}  & \includegraphics[height=0.65cm,width=0.63cm]{latex/Image/OMEGAPCSGBHOBAL.drawio.png}\\
    Full & \textcolor{blue}{fauldhouse}  &  \\
    \hline
\end{tabular}
\caption{\label{tab:example:postorder-alias}
Example of Pre-order alias (tables at the start) linearization with alias for "SELECT East Region FROM w WHERE East Region in ('fauldhouse', 'newtongrange') GROUP BY East Region ORDER BY COUNT ( id ) DESC LIMIT 1", where natural Language question ="'which team has made the roll of honour more times in the east region south division: fauldhouse or newtongrange ?'"
}
\end{table*}

\end{document}